\documentstyle[12pt]{article}
\newcommand{\bff}[1]{{\mbox{\boldmath $#1$}}}
\begin{document}
\title{\bf
Regular and chaotic dynamics in time-dependent 
relativistic mean-field theory}
\author{\\[2.0ex]
D. Vretenar\footnotemark[1], P. Ring, G.\,A. Lalazissis and W. P\" oschl\\
Physik-Department der Technischen Universit\"at M\"unchen,\\
D-85748 Garching, Germany}
\vspace{10mm}
\date{\today}
\maketitle
\footnotetext[1]{Alexander von Humboldt Fellow,
on leave of absence from University of Zagreb, Croatia}
\begin{abstract}
Isoscalar and isovector monopole oscillations that
correspond to giant resonances in spherical nuclei are
described in the framework of time-dependent relativistic
mean-field theory. Time-dependent and self-consistent 
calculations that reproduce experimental data on 
monopole resonances in $^{208}$Pb show that 
the motion of the collective coordinate is regular 
for isoscalar oscillations, and that it becomes chaotic
when initial conditions correspond to the isovector
mode. Regular collective dynamics coexists with 
chaotic oscillations on the microscopic level.
Time histories, Fourier spectra, state-space plots,
Poincar\' e sections, autocorrelation functions, and
Lyapunov exponents are used to characterize the 
nonlinear system and to identify chaotic oscillations.
Analogous considerations apply to higher multipolarities.
\end{abstract}

\newpage
\baselineskip = 24pt

\section{Introduction}

In the last two decades many studies have been reported 
in which the atomic nucleus has been used as a laboratory,
both experimentally and theoretically, for investigating
the transition from order to chaos in quantum dynamical
systems (for a recent review see~\cite{Zel.96}).  
Most of these studies have concentrated on two
major aspects: (i) generic signatures of chaos in 
local fluctuations and correlations of nuclear 
level distributions, and (ii) chaos in microscopic and
collective dynamics of realistic many-body systems.
In the first case, signatures of quantum chaos 
have also been studied in the complicated structure
of wave functions and randomness of matrix elements 
of physical operators. On the other hand, the nature
of collective nuclear dynamics has been investigated 
with particular emphasis on the stability of low-lying
nuclear modes in relation to one-body dissipation 
caused by the deformation of the nuclear potential,
cluster effects and Coriolis forces. The most elementary 
collective modes in nuclei are giant resonances. 
These are highly collective nuclear excitations in which a 
large fraction of nucleons participate. They can be described
as damped harmonic/anharmonic density oscillations 
around the equilibrium ground-state. Giant resonances
occur over the whole periodic table and their 
characteristic parameters are smooth functions
of the mass number. A mean-field model therefore
provides an appropriate framework for the 
description of giant resonances. Regular and chaotic 
dynamics in giant nuclear oscillations has been the subject 
of a number of studies. What has emerged as a very 
interesting result is that an undamped collective mode 
may coexist with chaotic single-particle motion. 
It appears that the slowly vibrating self-consistent
mean field created by the nucleons averages 
out the random components in their motion.
In all investigations the motion of 
only one type of particles has been considered.
That is, only the dynamics of isoscalar 
collective modes.

In the present article we study the difference in the 
dynamics of isoscalar and isovector collective modes. 
In particular, we describe isoscalar and isovector
monopole oscillations in spherical nuclei, but 
analogous considerations apply to higher multipolarities.
The dynamics of nuclear vibrations is analyzed in the framework
of time-dependent relativistic mean-field theory (TD RMFT).
The model represents a relativistic generalization of the
time-dependent Hartree-Fock approach.  Nuclear dynamics is
described by the simultaneous evolution of $A$ single
particle Dirac spinors in the time-dependent mean fields.
Frequencies of eigenmodes are determined from a Fourier
analysis of dynamical quantities. In this microscopic
description, self-consistent mean-field calculations are
performed for static ground-state properties, and
time-dependent calculations for monopole excitations. 
Because the time-evolution is 
calculated self-consistently, the system is 
intrinsically non-linear.
An advantage of the time-dependent approach is that no assumption
about the nature of the mode of vibrations has to be made.

The article is organized as follows. In Sec.~2 we present
the essential features of the time-dependent relativistic
mean-field model, as well as some  details of its application to
spherical nuclei. Time-dependent calculations
of isoscalar and isovector monopole oscillations 
in $^{16}$O and $^{208}$Pb are described in Sec.~3.
Results of a number of diagnostic tests that are used  to
identify chaotic oscillations in the nuclear system are 
discussed and compared with earlier work.
A summary of our results is presented in Sec.~4.

\section{Outline of the model}

The dynamics of collective vibrations in spherical nuclei 
will be described in
the framework of relativistic mean-field
theory~\cite{SW.86,Ser.92,Rin.96}.
In relativistic quantum hadrodynamics
the nucleus is described as a system of Dirac nucleons which
interact through the
exchange of virtual mesons and photons.  The Lagrangian
density of the model is
\begin{eqnarray}
{\cal L}&=&\bar\psi\left(i\gamma\cdot\partial-m\right)\psi
~+~\frac{1}{2}(\partial\sigma)^2-U(\sigma )
\nonumber\\
&&-~\frac{1}{4}\Omega_{\mu\nu}\Omega^{\mu\nu}
+\frac{1}{2}m^2_\omega\omega^2
~-~\frac{1}{4}{\vec{\rm R}}_{\mu\nu}{\vec{\rm R}}^{\mu\nu}
+\frac{1}{2}m^2_\rho\vec\rho^{\,2}
~-~\frac{1}{4}{\rm F}_{\mu\nu}{\rm F}^{\mu\nu}
\nonumber\\
&&-~g_\sigma\bar\psi\sigma\psi~-~
g_\omega\bar\psi\gamma\cdot\omega\psi~-~
g_\rho\bar\psi\gamma\cdot\vec\rho\vec\tau\psi~-~
e\bar\psi\gamma\cdot A \frac{(1-\tau_3)}{2}\psi\;.
\label{lagrangian}
\end{eqnarray}
The Dirac spinor $\psi$ denotes the nucleon with mass $m$.
$m_\sigma$, $m_\omega$, and $m_\rho$ are the masses of the
$\sigma$-meson, the $\omega$-meson, and the $\rho$-meson,
and $g_\sigma$, $g_\omega$, and $g_\rho$ are the
corresponding coupling constants for the mesons to the
nucleon. $U(\sigma)$ denotes the nonlinear $\sigma$
self-interaction
\begin{equation}
U(\sigma)~=~\frac{1}{2}m^2_\sigma\sigma^2+\frac{1}{3}g_2\sigma^3+
\frac{1}{4}g_3\sigma^4,
\label{NL}
\end{equation}
and $\Omega^{\mu\nu}$, $\vec R^{\mu\nu}$, and $F^{\mu\nu}$
are field tensors~\cite{SW.86}.

The coupled equations of motion are derived from the Lagrangian
density (\ref{lagrangian}).
The Dirac equation for the nucleons:
\begin{eqnarray}
i\partial_t\psi_i&=&\left[ \bff\alpha
\left(-i\bff\nabla-g_\omega\bff\omega-
g_\rho\vec\tau\vec{\bff\rho}
-e\frac{(1-\tau_3)}{2}{\bff A}\right)
+\beta(m+g_\sigma \sigma)\right. \nonumber\\
&&\left. +g_\omega \omega_0+g_\rho\vec\tau\vec\rho_0
+e\frac{(1-\tau_3)}{2} A_0
\right]\psi_i
\label{dirac}
\end{eqnarray}
and the Klein-Gordon equations for the mesons:
\begin{eqnarray}
\left(\partial_t^2-\Delta+m^2_\sigma\right)\sigma&=&
-g_\sigma\rho_s-g_2 \sigma^2-g_3 \sigma^3\\
\left(\partial_t^2-\Delta+m^2_\omega\right)\omega_\mu&=&
~g_\omega j_\mu\\
\left(\partial_t^2-\Delta+m^2_\rho\right)\vec\rho_\mu&=&
~g_\rho \vec j_\mu\\
\left(\partial_t^2-\Delta\right)A_\mu&=&
~e j_\mu^{\rm em}.
\label{KGeq4}
\end{eqnarray}
In the relativistic mean field approximation the A nucleons
described by single-particle spinors $\psi_i,~(i=1,2,...,A)$, are
assumed to form the A-particle Slater determinant $|\Phi\rangle$,
and to move independently in the classical meson fields.
The sources of the fields, i.e.
densities and currents, are calculated in the {\it no-sea}
approximation~\cite{VBR.95}:\ -~the scalar density
\begin{equation}
\rho_{\rm s}~=~\sum_{i=1}^A \bar\psi_i\psi_i,
\label{rho}
\end{equation}
-~the isoscalar baryon current
\begin{equation}
j^\mu~=~\sum_{i=1}^A \bar\psi_i\gamma^\mu\psi_i,
\label{current}
\end{equation}
-~the isovector baryon current
\begin{equation}
\vec j^{\,\mu}~=~\sum_{i=1}^A \bar\psi_i\gamma^\mu \vec \tau\psi_i,
\label{isocurrent}
\end{equation}
-~the electromagnetic current for the photon-field
\begin{equation}
j^\mu_{\rm em}~=~\sum_{i=1}^A
\bar\psi_i\gamma^\mu\frac{1-\tau_3}{2}\psi_i,
\label{emcurrent}
\end{equation}
where the summation is over all occupied states in the
Slater determinant $|\Phi\rangle$. Negative-energy states
do not contribute to the densities in the {\it no-sea}
approximation of the stationary solutions.  However,
negative energy contributions are included implicitly in
the time-dependent calculation, since the Dirac equation is
solved at each step in time for a 
different basis set~\cite{VBR.93,VBR.95}.
Negative energy components with respect to the original
ground state basis are taken into account automatically,
even if at each time the {\it no-sea} approximation is
applied.  It is also assumed that nucleon single-particle
states do not mix isospin. 

The ground state of a nucleus is described by
the stationary self-consistent solution of the
coupled system of equations
(\ref{dirac})--(\ref{KGeq4}),
for a given number of nucleons
and a set of coupling constants and masses.
The solution for the ground state specifies part of the initial
conditions for the time-dependent problem.
For a given set of initial conditions, i.e. initial values
for the densities and currents in Eqs.
(\ref{rho}--\ref{emcurrent}), the model describes the time
evolution of $A$ single-particle wave functions in the
time-dependent mean fields.
Retardation effects for the meson fields are not
included, i.e. the time derivatives
$\partial_t^2$ in the equations of motions for the meson
fields are neglected. This is justified by the large masses 
in the meson propagators, causing a short range
of the corresponding meson exchange forces. 
Since there is no systematic
procedure to derive the initial conditions that
characterize the motion of a specific mode of the nuclear
system, the description of nuclear dynamics as a
time-dependent initial-value problem is intrinsically
semi-classical. The theory can be quantized by the
requirement that there exist time-periodic solutions of the
equations of motion, which give integer multiples of
Planck's constant for the classical action along one period
~\cite{RVP.96}.  For giant resonances the time-dependence
of collective dynamical quantities is actually not
periodic, since generally giant resonances are not
stationary states of the mean-field Hamiltonian. The
coupling of the mean-field to the particle continuum allows
for the decay of giant resonances by direct escape of
particles.  In the limit of small amplitude oscillations,
however, the energy obtained from the frequency of the
oscillation coincides with the excitation energy of the
collective state. In Refs.~\cite{VBR.95,PVR.96,RVP.96,Vre.97} we have shown that the model reproduces reasonably well
the experimental data on giant resonances in spherical nuclei.

In the present article we apply the model to
isoscalar and isovector monopole oscillations in spherical
nuclei.  In this microscopic description, self-consistent
mean-field calculations are performed for static
ground-state properties, and time-dependent calculations
for monopole excitations. Because of the 
self-consistent time-evolution, the system is 
intrinsically non-linear. For a system with spherical    
symmetry, the nucleon single-particle spinor
is characterized by the angular
momentum $j$, its $z$-projection $m$, the parity $\pi$, and
the isospin $t_3=\pm\frac{1}{2}$ for neutrons and
protons
\begin{equation}
\psi({\bf r},t,s,t_3) =
\frac{1}{r}
\left(
\begin{array}{c}
\ \ f(r) \Phi_{ljm} (\theta, \varphi, s) \\
  ig(r) \Phi_{\tilde ljm} (\theta, \varphi, s)
\end{array}
\right)
\, e^{-iEt} \, \chi_\tau(t_3).
\label{wavefunction}
\end{equation}
$\chi_\tau$ is the isospin function, the orbital angular
momenta $l$ and $\tilde l$ are determined by $j$ and the
parity $\pi$, $f(r)$ and $g(r)$ are radial functions, and
$\Phi_{ljm}$ is the tensor product of the orbital and spin
functions
\begin{equation}
\Phi_{ljm}(\theta, \varphi, s)~=~\sum_{m_s m_l} < \frac{1}{2}~m_s
~l~m_l | j~m> Y_{l m_l}(\theta, \varphi)~\chi_{m_s}(s).
\end{equation}
In order to excite monopole oscillations, the spherical solution 
for the ground-state has to be initially compressed or radially
expanded by scaling the radial coordinate. The amplitudes
$f^{\rm mon}$ and $g^{\rm mon}$ of the Dirac spinor are
defined
\begin{equation}
f^{\rm mon}(r_{\rm mon})  =
(1+a)~f(r),~~~
g^{\rm mon}(r_{\rm mon})  =
(1+a)~g(r)\;.
\label{def}
\end{equation}
The new coordinates are
\begin{equation}
r_{\rm mon}= (1+a)~r.
\end{equation}
For isoscalar oscillations the monopole
deformations of the proton and neutron densities have the
same sign. To excite isovector oscillations, the initial
monopole deformation parameters of protons and neutrons
must have opposite signs.  After the initial 
deformation (\ref{def}), the proton and neutron densities
have to be normalized.

The time-dependent
Dirac equation (\ref{dirac}) is reduced to a set of coupled
first-order partial differential equations for the complex
amplitudes $f$ and $g$ of proton and neutron states
\begin{eqnarray}
i\partial_t f&=&(V_0 + g_\sigma \sigma)f+
(\partial_{r}
- {\kappa\over r}- iV_r)g\\
i\partial_t g&=&(V_0-g_\sigma \sigma -2m)g-
(\partial_{r}
+ {\kappa\over r}- iV_r)f,
\end{eqnarray}
where $\kappa = \pm (j+ {1\over 2})$ for $j = l \mp {1\over
2}$, and the indices $0$ and $r$ denote the time and radial
components of the vector field
\begin{equation}
V^\mu= g_\omega \omega^\mu + g_\rho \rho_3^\mu \tau_3 +
e{(1-\tau_3)\over 2}A^\mu.
\end{equation}
For a given set of initial conditions, the equations of
motion propagate the nuclear system in time. The potentials
are solutions of the Klein-Gordon equations
\begin{equation}
\left(-{\partial^2 \over {\partial r^2}} -
{2\over r}{\partial \over {\partial r}}
+m^2_\phi\right) \phi(r)~=~s_\phi (r),
\end{equation}
$m_\phi$ are meson masses for $\phi = \sigma$, $\omega$ and
$\rho$, and zero for the photon.  The source terms are
calculated from~(\ref{rho})--(\ref{emcurrent}) using in
each time-step the latest values for the nucleon
amplitudes.  
The meson fields and electromagnetic potentials are
calculated from
\begin{equation}
\phi(r)~=~\int_0^\infty G_\phi (r,r^\prime) s_\phi(r^\prime)
r^{\prime 2} dr^\prime\;,
\end{equation}
where for massive fields
\begin{equation}
G_\phi(r,r^\prime)~=~{1\over {2m_\phi}} {1\over{r r^\prime}}
\left( e^{m_\phi |r-r^\prime|} - e^{-m_\phi|r+r^\prime|}\right),
\end{equation}
and for the Coulomb field
\begin{eqnarray}
G_C(r,r^\prime)&=&
\frac {1}{r}  ~~~~~\mbox{for}~~ r > r^\prime \nonumber\\
G_C(r,r^\prime)&=&
\frac{1}{r^\prime}  ~~~~~\mbox{for}~~ r <  r^\prime.
\end{eqnarray}
The collective dynamical variables that characterize vibrations of a
nucleus are defined as expectation values of
single-particle operators in the time-dependent Slater
determinant $|\Phi(t)\rangle$ of occupied states.  In the
framework of the TDRMF model the wave-function of the
nuclear system is a Slater determinant at all times.  For
isoscalar monopole vibrations, the time-dependent monopole
moment is defined:
\begin{equation}
\langle r^2(t)\rangle ~=~\frac {1}{A}\langle \Phi(t) |r^2 |\Phi(t)\rangle.
\end{equation}
The corresponding isovector monopole moment is simply
\begin{displaymath}
< r^2_{\rm p}(t) > - < r^2_{\rm n} (t)>.
\end{displaymath}
Fourier transforms of the collective dynamical variable determine the
frequencies of eigenmodes.  

\section{Isoscalar and isovector monopole oscillations }

The study of isoscalar monopole resonances in nuclei
provides an important source of information on the nuclear
matter compression modulus $K_{\rm nm}$. This quantity is
crucial in the description of properties of
nuclei, supernovae explosions, neutron stars, and heavy ion
collisions. Modern non-relativistic
Hartree-Fock plus RPA calculations, using both Skyrme and
Gogny effective interactions, indicate that the value of
$K_{\rm nm}$ should be in the range 210-220 MeV~\cite{BBD.95a}.
In the framework of relativistic mean field theory on the other hand, 
calculations based on an effective interaction \cite{LKR.97} with 
nuclear matter compression modulus $K_{\rm nm}\approx 250
- 270$ MeV are in better agreement with available
data on spherical nuclei~\cite{Vre.97}.
The complete experimental data set on
isoscalar giant monopole resonances (GMR) has been recently
analyzed by Shlomo and Youngblood~\cite{SY.93}.
In Fig. 1 we display results of time-dependent 
relativistic mean-field calculations of isoscalar
and isovector oscillations in $^{208}$Pb. 
The experimental isoscalar GMR energy in $^{208}$Pb is 
well established at $13.7\pm 0.3$ MeV. 
Experimental data on isovector giant monopole resonances
are much less known. The systematics of excitation energies
does not, in general, depend on the nuclear matter
compression modulus. 
The experimental value for the isovector GMR in $^{208}$Pb:
$26\pm 3$ MeV~\cite{GBM.90}. We have calculated 
the ground-state with the NL1~\cite{RRM.86}
parameter set of the effective Lagrangian. This 
effective force has been extensively
used in the description of properties of finite nuclei
along the valley of beta stability~\cite{Rin.96}.
For NL1 ($K_{\rm nm} =211.7$ MeV), we expect 
the calculated excitation
energy for the isoscalar mode to be
approximately 1-2 MeV lower than the average experimental
value. However, the precise value is not important 
in the present consideration.
For the initial deformation parameter in (\ref{def}) we 
have used $a=0.2$. In the isoscalar case both proton and 
neutron densities are radially expanded, while for the
isovector mode the proton density is initially compressed
by the same amount. Therefore although the initial
conditions are different, in both cases we follow
the time evolution of the same system.

In Fig. 1a we plot the time history of the isoscalar 
monopole moment $\langle r^2(t)\rangle$, and in Fig. 1b
the corresponding isovector moment is shown. The 
isoscalar mode displays regular undamped oscillations, 
while for the isovector mode we observe strongly 
damped anharmonic oscillations. On the right hand panels we
plot the corresponding Fourier power spectra. 
The evolution time $T_{\rm final}= 3000~{\rm fm/c}$
determines the numerical resolution $\Delta E = 2\pi
\hbar c / T_{\rm final} = \approx 0.4$ MeV in the frequency
domain. The Fourier power spectra are calculated by FFT
using data windowing with the Welch window function. We
have verified that the positions of the main peaks do not
depend on the choice of the window function. We estimate
the numerical accuracy to $\approx 0.5$ MeV.
As one would expect for a heavy
nucleus, there is very little spectral fragmentation 
in the isoscalar channel, and a
single mode dominates at the excitation energy of 11 MeV.
The Fourier spectrum of the isovector mode is strongly fragmented. 
However, the main peaks are found in the energy region $25 - 30$ MeV, 
in agreement with the experimental data. 

For the isoscalar mode, the time history of the monopole moment 
and the Fourier spectrum show that the oscillations of the collective 
coordinate are regular. On the other hand, the appearance of a 
broad spectrum of frequencies seems 
to indicate that the isovector oscillations are chaotic.
A diagnosis of chaotic vibrations would imply that one has
a clear definition of such motion. For a quantum system
however, the concept of chaos, especially in time-dependent
problems, is not well defined. And although our description 
of nuclear vibrations is semi-classical, quantum effects
like the Pauli principle are present in the initial 
conditions and during the dynamical evolution. 
A number of diagnostic tests can help to 
identify chaotic oscillations in physical systems~\cite{NB.95},
and some of them can be applied in the present consideration.
In Figs. 2 - 4 we display some additional qualitative measures 
which can be used to characterize the response of our 
nonlinear system. In Fig. 2 we have constructed the two-dimensional 
time-delayed pseudo-phase space for isoscalar (a), and
isovector (b) oscillations shown in Fig. 1. Since information
is available on the time evolution of just one variable, the 
collective coordinate, one plots the signal versus itself, but
delayed or advanced by a fixed time constant $[<r^2(t)>, 
<r^2(t+\tau)>]$. The idea is that the signal $<r^2(t+\tau)>$ 
is related to $<\dot{r}^2(t)>$,
and should have properties similar to those in the classical 
phase plane $[<r^2(t)>, <\dot{r}^2(t)>]$. The choice of the time delay
$\tau$ is not crucial, except to avoid a natural period of the system. 
If the motion is chaotic, the trajectories in the phase space 
do not close. For the pseudo-phase planes shown in Fig. 2, we 
have taken $\tau = 20~fm/c$. The phase space trajectories for 
the isoscalar mode are closed ellipses, indicating regular oscillations.
For the isovector oscillations on the other hand, the trajectories
are completely chaotic. The strong damping results from 
one-body processes: (i) escape of nucleons into the continuum
states and (ii) collisions of the 
nucleons with the moving wall of the nuclear potential 
generated by the self-consistent mean fields.
In Fig. 3 we display the corresponding Poincar\' e sections 
constructed from three-dimensional time-delayed pseudo-phase space
$[<r^2(t)>, <r^2(t+\tau)>, <r^2(t+ 2 \tau)>]$, with $\tau = 20~fm/c$. 
The Poincar\' e maps are shown in the planes: 
$<r^2(t)> = 35~fm^2$ for the isoscalar mode, and $<r^2(t)> = -9.55~fm^2$ 
for the isovector mode, respectively. The Poincar\' e map
for the isoscalar mode consists of two sets
of closely located points and therefore confirms regular oscillations.
For the isovector oscillations the Poincar\' e map appears as 
a cloud of unorganized points in the phase plane. Such a map 
indicates stochastic motion. Another measure that is related to 
the Fourier transform is the autocorrelation function 
\begin{equation}
A(\tau) = \lim_{T\to \infty} \int\limits_{0}^{T}
           <r^2(t)> <r^2(t+\tau)>\,dt
\end{equation}
When the signal is chaotic, information about its past origins 
is lost. This means that $A(\tau) \to 0$ as $\tau \to \infty$, 
or the signal is only correlated with its recent past. It is also
expected $A(\tau)$ of a chaotically modulated signal to be an 
irregularly modulated waveform.
The autocorrelation functions for isoscalar and isovector oscillations
are shown in Fig. 4. The normalization is $A(\tau = 0) = 1$. 
For the isovector mode $A(\tau)$ shows a rapid decrease and the 
envelope appears as an 
irregular waveform. Therefore also this quantity
indicates that the dynamical variable displays chaotic
oscillations for the isovector mode. 

In our first example we have considered monopole oscillations 
in $^{208}$Pb, for which there exist experimental data on 
both isoscalar and isovector giant monopole resonances. 
Thus our description of the dynamics of monopole oscillations
is not just a completely artificial model, 
but in fact corresponds to an experimentally observed 
physical system. On the other hand $^{208}$Pb is a large 
and complicated system, in which many single-nucleon 
orbitals contribute to the collective coordinate. 
In order to study in more detail the dynamics, 
it is more convenient to 
consider a light nucleus: $^{16}$O. In Fig. 5 
we compare the isoscalar and isovector monopole moments 
and the associated Fourier power spectra. For the isoscalar
mode a modulation of the signal is observed, and
two strong peaks at approximately 20 MeV excitation 
energy are found in the Fourier spectrum. 
A fragmentation in the Fourier spectrum 
is expected for such a light nucleus. The behavior of
the collective coordinate for the isovector mode appears 
to be chaotic. The Fourier spectrum is strongly fragmented 
in the energy region $15 - 45$ MeV. The pseudo-phase spaces 
are compared in Fig. 6.
The results are similar to those for $^{208}$Pb and
indicate that isovector oscillations are chaotic.

In Refs.~\cite{Blo.92,Blo.93} Blocki {\it et al.} analyzed 
the behavior of a purely classical gas of noninteracting point
particles enclosed in a hard-wall container, which undergoes 
periodic and adiabatic shape oscillations. The wall motion is 
not modified by the collisions with the particles, i.e. the 
dynamics is not self-consistent. Analyses of Poincar\' e
sections and of Lyapunov exponents showed that for low
deformations ($l=2$) the motion of particles is regular, 
whereas for higher multipolarities ($l\geq 3$) the scattering 
of segments of wall with positive curvature leads to 
divergence between trajectories and therefore to chaotic 
motion. A somewhat different result, especially if applied 
to the nuclear system, was found in Ref.~\cite{Bau.94}. 
Conditions under which nucleons inside a 
nucleus can undergo chaotic motion were studied. A self-consistent
model of separable forces was used in a semi classical 
approximation of the time-dependent Hartree-Fock 
equation. The test particle method was used to solve the 
Vlasov equation for the time-evolution of the density
matrix. Isoscalar quadrupole and octupole oscillations 
were investigated. It was shown that, both for quadrupole
and octupole deformations, chaotic single-particle
dynamics leads to regular motion of the collective 
coordinate, the multipole moment of deformation. The 
origin of the chaotic single-particle dynamics was 
attributed to the exchange of energy between the motion
of the individual test particles and the collective motion 
of the multipole coordinate. In particular, it was stressed
that self-consistency is essential for the generation 
of regular dynamics from an ensemble of single-nucleons 
with chaotic trajectories. The relationship between 
chaoticity at the microscopic level and dissipation 
of the collective degrees of freedom was also investigated 
in Ref.~\cite{Bur.95}. Classical particles were confined 
in a two-dimensional nuclear potential whose walls 
undergo periodic shape oscillations of fixed multipolarity.
The collective variable appears explicitly in the 
Hamiltonian as an additional degree of freedom. A fully 
self-consistent description of the dynamics of particle motion 
and the collective coordinate was employed. Isoscalar 
monopole oscillations were studied and it was shown 
that self-consistency induces chaotic single-particle motion.
In particular, it was demonstrated how the coupling between
the collective variable and single-particle dynamics 
induces macroscopic dissipation.

In a quantum description of nuclear dynamics, such as the 
one based on time-dependent relativistic mean-field theory,
we cannot follow the trajectories of individual 
nucleons. In order to better understand the difference 
between the isoscalar and isovector modes at the microscopic 
level, we consider the individual contributions of 
single-nucleon orbitals to the collective coordinate, the
time-dependent monopole moment. In the case of $^{16}$O 
($Z = N = 8$), protons and neutrons occupy 
only three spherical levels in the ground state:
$1s~1/2$, $1p~3/2$, and $1p~1/2$. 
Monopole oscillations do not break the degeneracy, i.e. 
spherical symmetry is conserved during the time evolution
of the nuclear system and we can follow the dynamics of 
each level. In Fig. 7 we display the time histories
of the expectation values of $r^2$ for the three neutron 
levels. In Fig. 7a we plot 
$<r^2 (t)>_{s 1/2, p 3/2, p 1/2}$ for the isoscalar mode, 
for which the oscillations of the collective coordinate 
are shown in Fig. 5a. Fig. 7b corresponds to the isovector 
monopole oscillations displayed in Fig. 5b. 

In contrast to the collective isoscalar and isovector 
moments shown in Fig. 5,
in Fig. 7a and 7b we plot the same quantities, expectation
values of $r^2$ for the neutron single-particle orbitals. 
Also the initial conditions for neutron motion are the 
same in both cases. The only difference is the initial 
condition for protons. In the isoscalar case
protons oscillate in phase with neutrons. Proton and 
neutron oscillations are out of phase for the isovector 
mode. In Fig. 7a one observes regular modulated oscillations
for all three neutron levels. For the isovector case the time
histories display chaotic oscillations. In Fig. 8 we display 
the corresponding Fourier power spectra and
the largest Lyapunov exponents. Comparing the Fourier spectra,  
we notice that the differences 
between isoscalar and isovector oscillations are especially
pronounced for the $1s~1/2$ level. Much more fragmentation 
is observed in the isovector spectrum, and the strength is
shifted toward higher frequencies. The
Fourier spectra for isovector oscillations display 
pronounced peaks in the higher frequency 
region. From the time series of the expectation 
values $<r^2 (t)>_{s 1/2, p 3/2, p 1/2}$, 
we have also calculated the largest Lyapunov exponents,  
shown in the right hand panels of Fig. 8.
Lyapunov exponents 
provide a qualitative and quantitative characterization 
of dynamical behavior. They are related to the exponentially 
fast divergence or convergence of nearby orbits in phase space.
A system with one or more positive Lyapunov exponents is defined
to be chaotic. The magnitude of the exponents reflects the time 
scale on which system dynamics becomes unpredictable.
The largest exponents are displayed as function of 
evolution time in Fig. 9. They
have been calculated by the method of Ref.~\cite{Wol.85}, 
which allows the estimation of non-negative Lyapunov exponents 
from a time series.
For the $1s~1/2$ level the calculated exponent for oscillations that
correspond to the isoscalar collective mode is consistent with 
the value zero, indicating regular motion. For the isovector 
mode, the Lyapunov exponent is positive and large. Therefore, 
oscillations of $<r^2 (t)>_{s 1/2}$ are chaotic for collective
isovector vibrations. For the $1p~3/2$ and $1p~1/2$ levels 
all calculated exponents are positive. For the isoscalar 
mode the small positive values reflect the observed fragmentation 
in the Fourier spectrum. Those that 
correspond to isovector collective mode are much larger and 
indicate that the dynamics is indeed chaotic. The corresponding Fourier 
spectra and largest Lyapunov exponents for proton levels are very similar. 
The only difference is that, in addition to meson exchange forces, 
the protons also feel the long range Coulomb interaction.

Similar results are obtained for $^{208}$Pb. We have calculated
the Fourier spectra and largest Lyapunov
exponents for the time series of expectation values $<r^2>$ for
the 22 single-particle spherical shell-model neutron states. 
Typical results are shown in Fig. 9 for the orbitals 
$1h~11/2$, $1h~9/2$, and $2f~7/2$. 
The column on the left corresponds to isoscalar monopole
collective oscillations, on the central panels the results for the 
isovector collective mode are shown. Both for isoscalar and 
isovector oscillations (Fig. 1), the initial conditions
for neutron motion are the same. All calculated Lyapunov 
exponents are positive, and therefore seem to indicate that 
the underlying microscopic dynamics is chaotic. In general, we 
observe much more fragmentation in the Fourier spectra of 
oscillations that correspond to the isovector mode, and 
the resulting Lyapunov exponents have also larger values. 
The interesting result of course is that for the isoscalar case we
observe regular oscillations of the collective variable. 
This would be in agreement with the results reported in 
Ref.~\cite{Bau.94}, where chaotic single-particle motion
was found in coexistence with regular collective dynamics.
For the isovector mode on the other hand, the collective
monopole moment displays chaotic oscillations. This is 
explained by the fact that protons and neutrons effectively 
move in two self-consistent potentials that oscillate out 
of phase. For example, when neutrons move inward, they 
scatter on the potential wall with positive curvature 
that is created by protons moving outward. This will 
lead to pseudo-random motion of the nucleons and 
dissipation of collective oscillations.
\section{Conclusions}
Isoscalar and isovector collective monopole oscillations in 
spherical nuclei have been analyzed within the framework
of time-dependent relativistic mean field theory.
These oscillations correspond to the most elementary collective
nuclear modes, the giant resonances. In order to investigate 
the dynamics of collective vibrations, we have analyzed 
time-dependent and self-consistent calculations that reproduce
the experimental data on monopole giant resonances in 
spherical nuclei. Because of the self-consistent time-evolution,  
the nuclear system is non-linear and one
expects chaotic dynamics for specific initial conditions. 
In particular, we have studied the difference in the 
dynamics of isoscalar and isovector collective modes. 
Time histories, Fourier spectra, state-space plots,
Poincar\' e sections, autocorrelation functions, and
Lyapunov exponents have been used to characterize the 
nonlinear system and to identify chaotic oscillations.
It has been shown that the oscillations of the collective
coordinate can be characterized as regular for the 
isoscalar mode, and that they become chaotic when
initial conditions correspond to the isovector mode. 
Our results also confirm the conclusions of a number 
of studies, which have shown how a regular collective mode 
can coexist with chaotic single-particle dynamics.  
However, we have shown that this is the case only 
for isoscalar modes, that is, only if one considers
the motion of a single type of particles. When 
protons and neutrons move out of phase, as it happens
for isovector modes, the resulting dynamics of the collective 
coordinate exhibit chaotic behavior. Of course, analogous considerations
apply also for higher multipolarities, for example
isoscalar and isovector giant quadrupole resonances. 
Due to numerical problems, for quadrupole oscillations 
our computer codes could not propagate the nuclear system
to very long times. However, a preliminary analysis of 
isoscalar and isovector quadrupole oscillations in 
$^{16}$O ($~T_{final} = 500~fm/c$) have shown that
results similar to the monopole resonances could be 
expected. Work along these lines is in progress.
\newpage
\vspace{20mm}
{\bf Acknowledgments}
\vspace{10mm}

This work has been supported in part by the
Bundesministerium f\"ur Bildung und Forschung under
contract 06~TM~875. 


\newpage

\leftline{\Large {\bf Figure Captions}}
\parindent = 2 true cm
\begin{description}

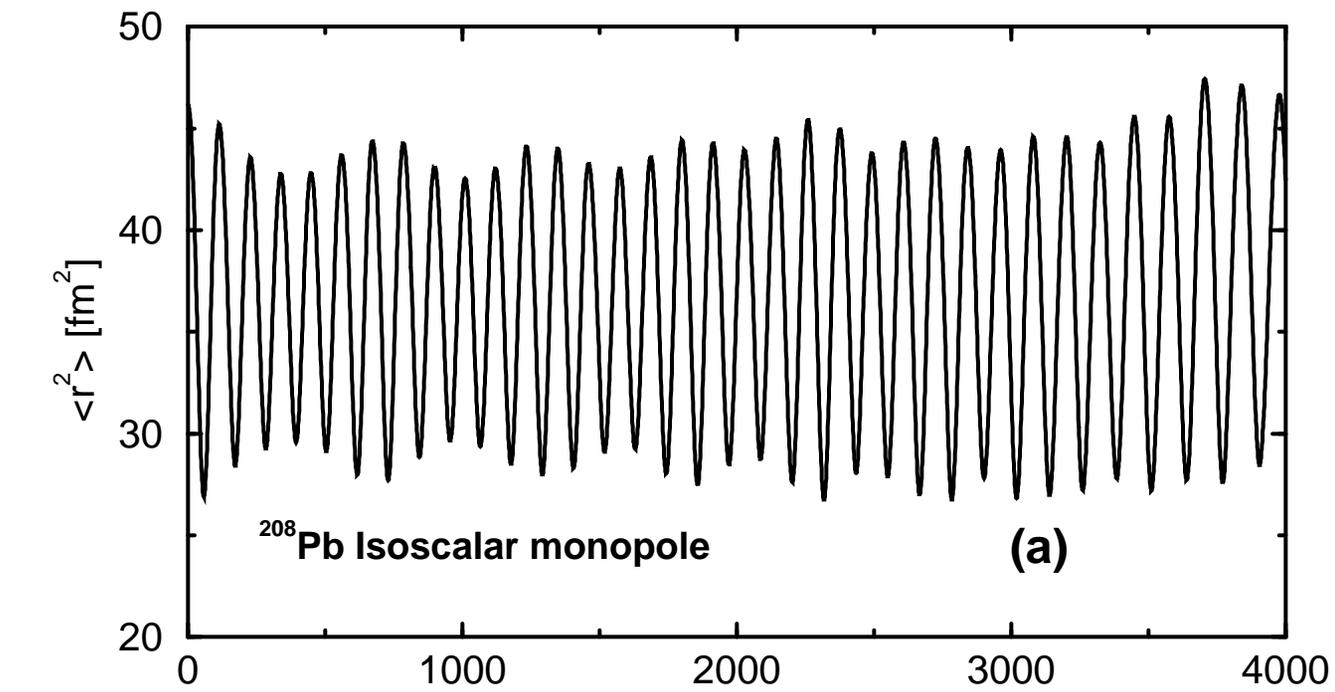
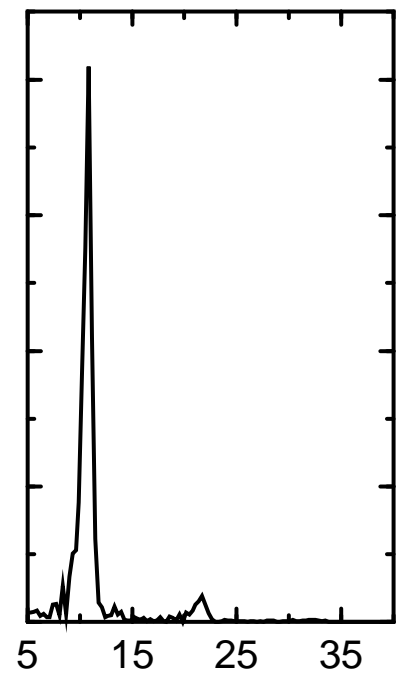
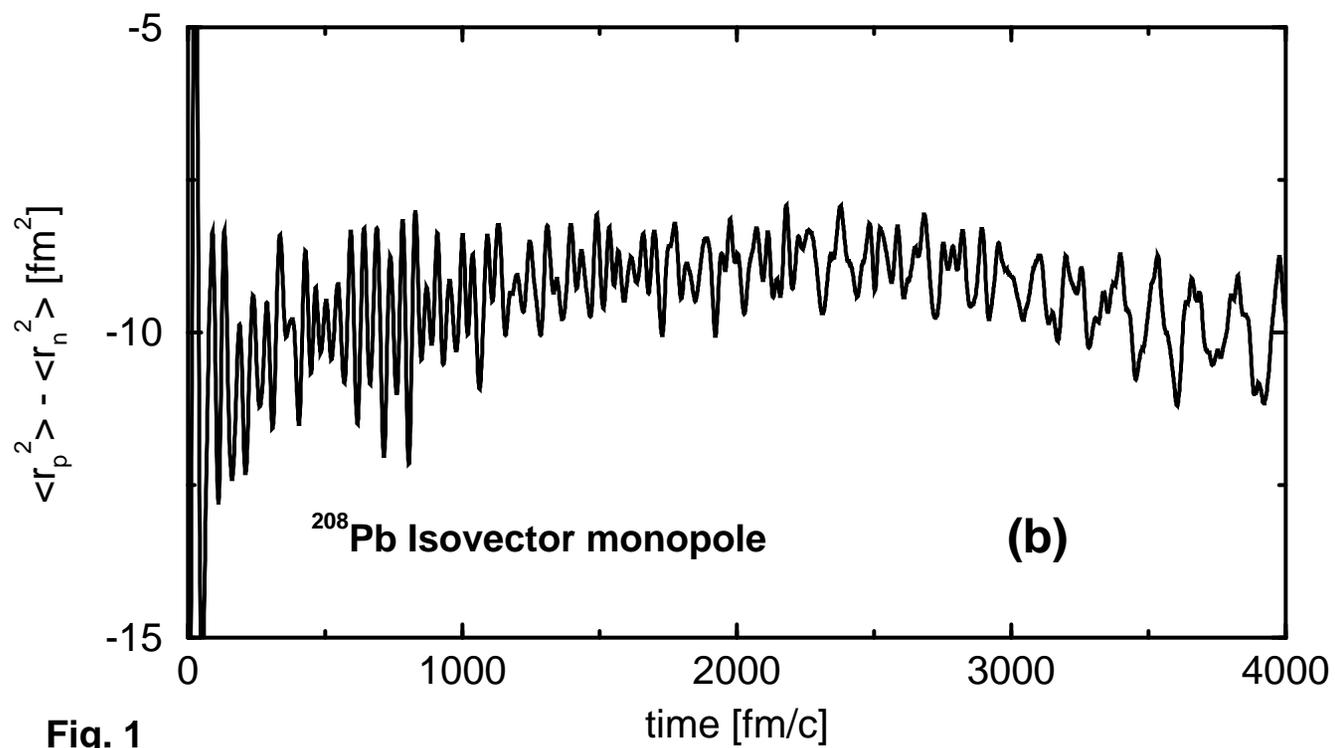
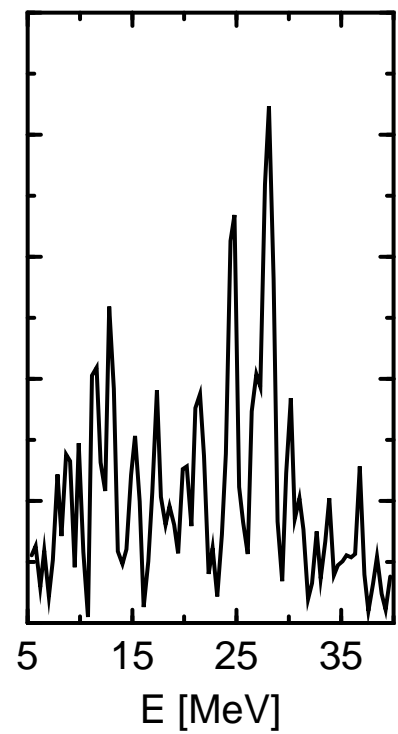
\item[Fig.~1] Time-dependent isoscalar $<r^2>$ (a) and 
isovector $<r_{\rm p}^2> - <r_{\rm n}^2>$ (b) monopole moments
and the corresponding Fourier power spectra for
$^{208}$Pb. The parameter set of the effective Lagrangian 
is NL1. 

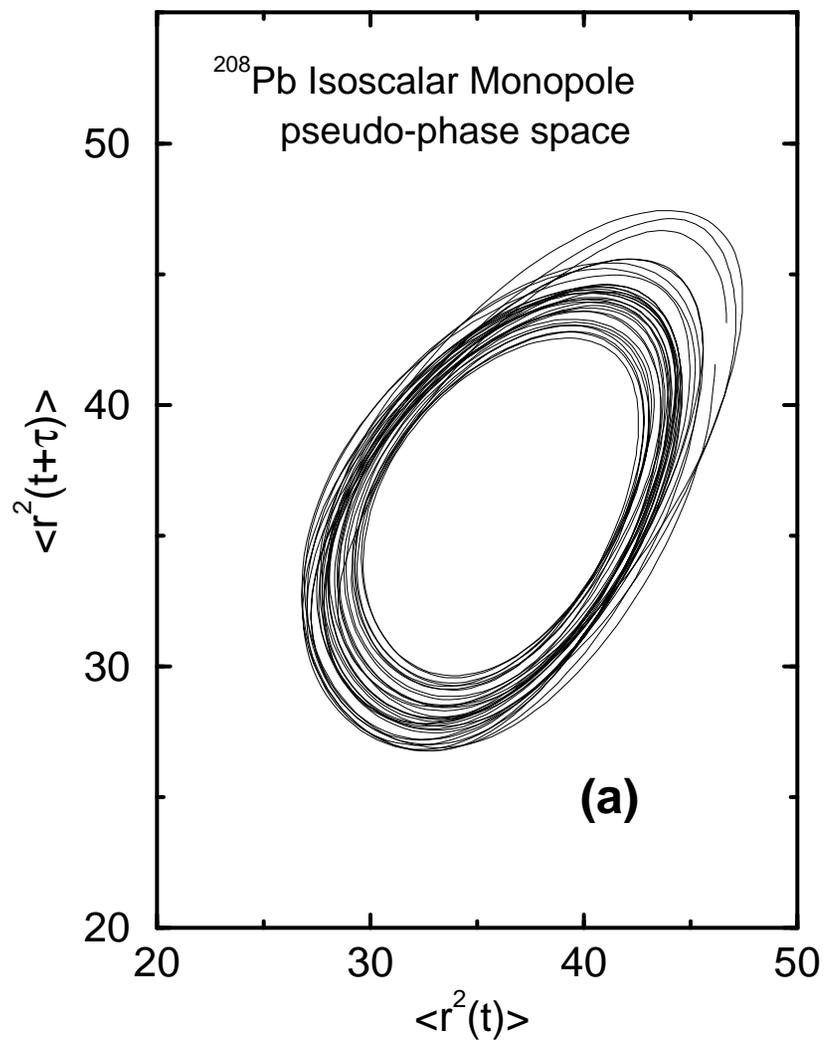
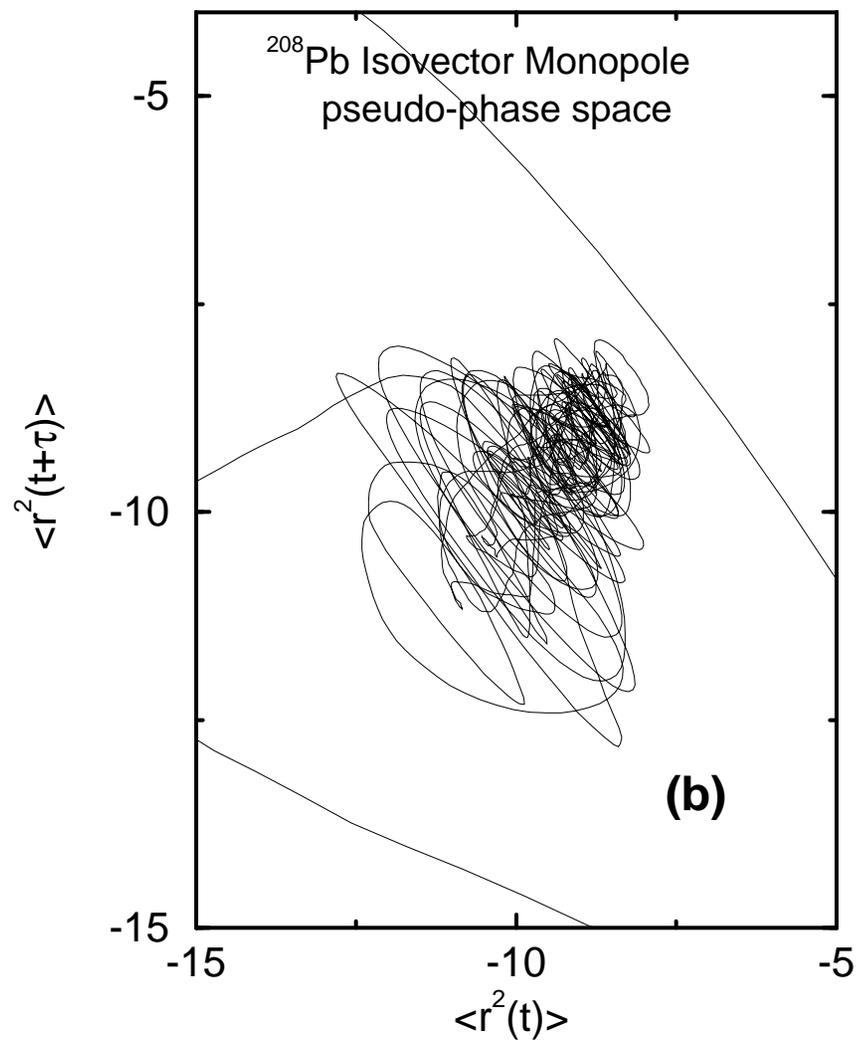
\item[Fig.~2] Two-dimensional pseudo-phase space for 
isoscalar (a) and isovector (b) monopole oscillations 
in $^{208}$Pb.

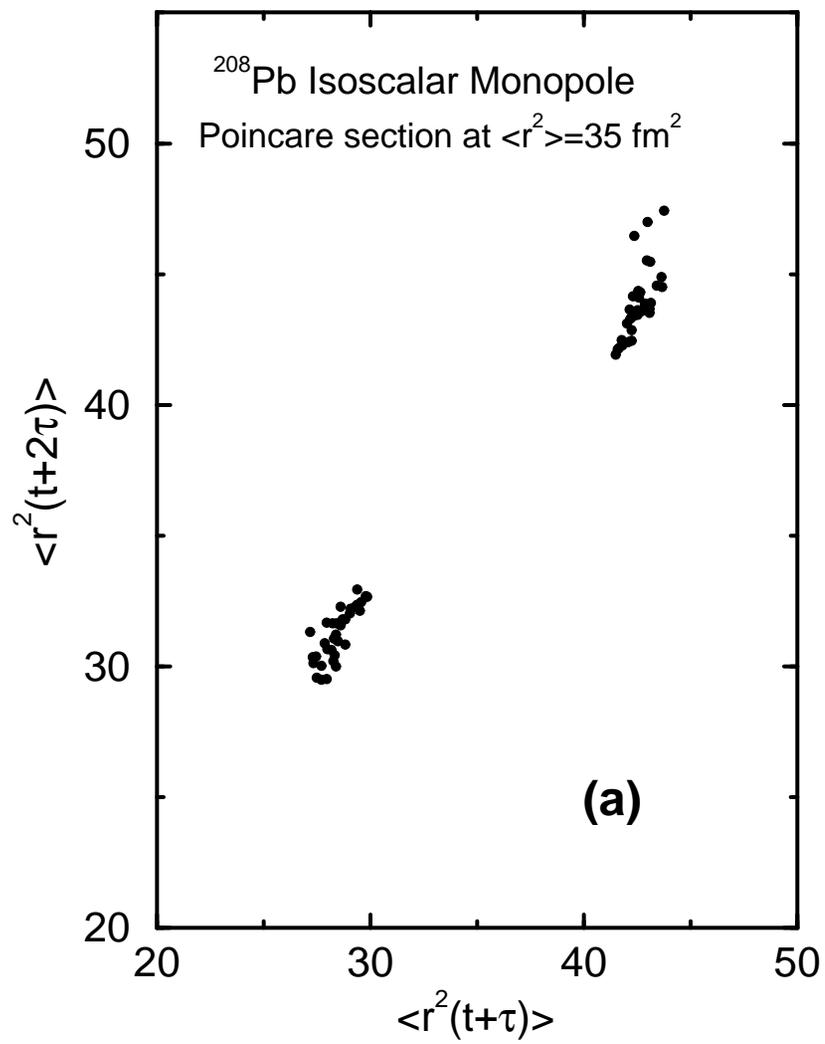
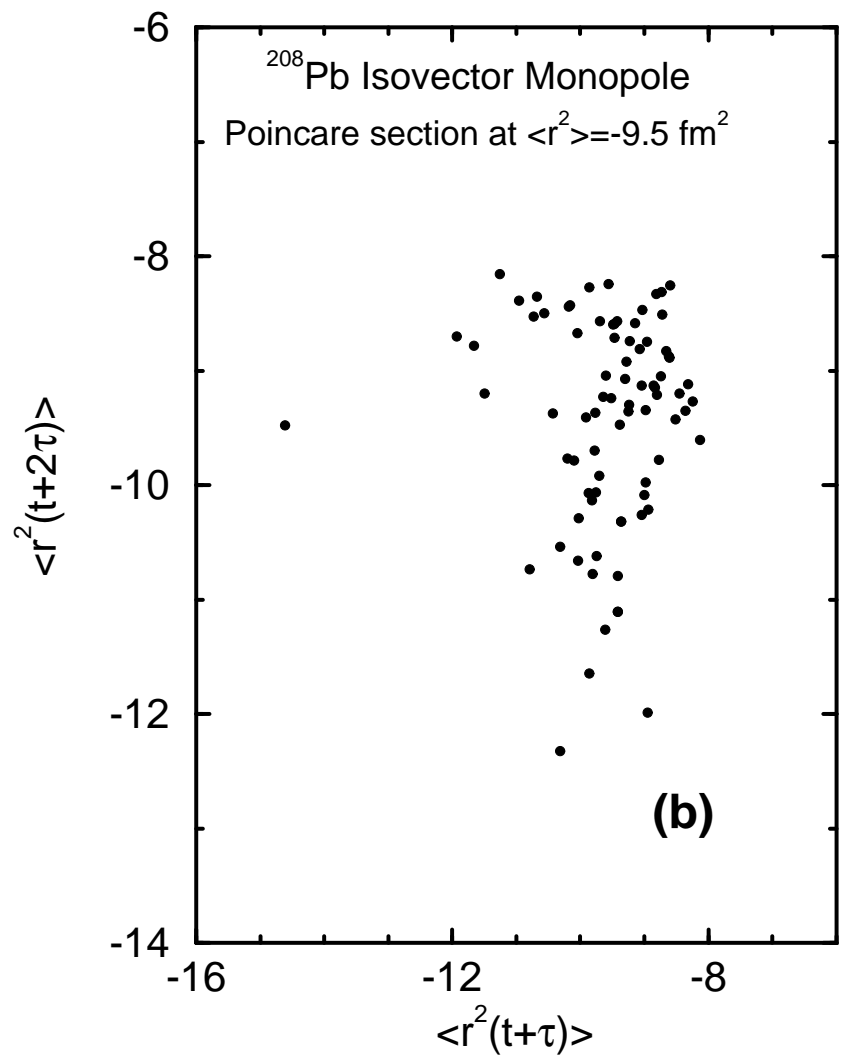
\item[Fig.~3] Poincar\' e sections for isoscalar (a) 
and isovector (b) monopole oscillations 
in $^{208}$Pb.

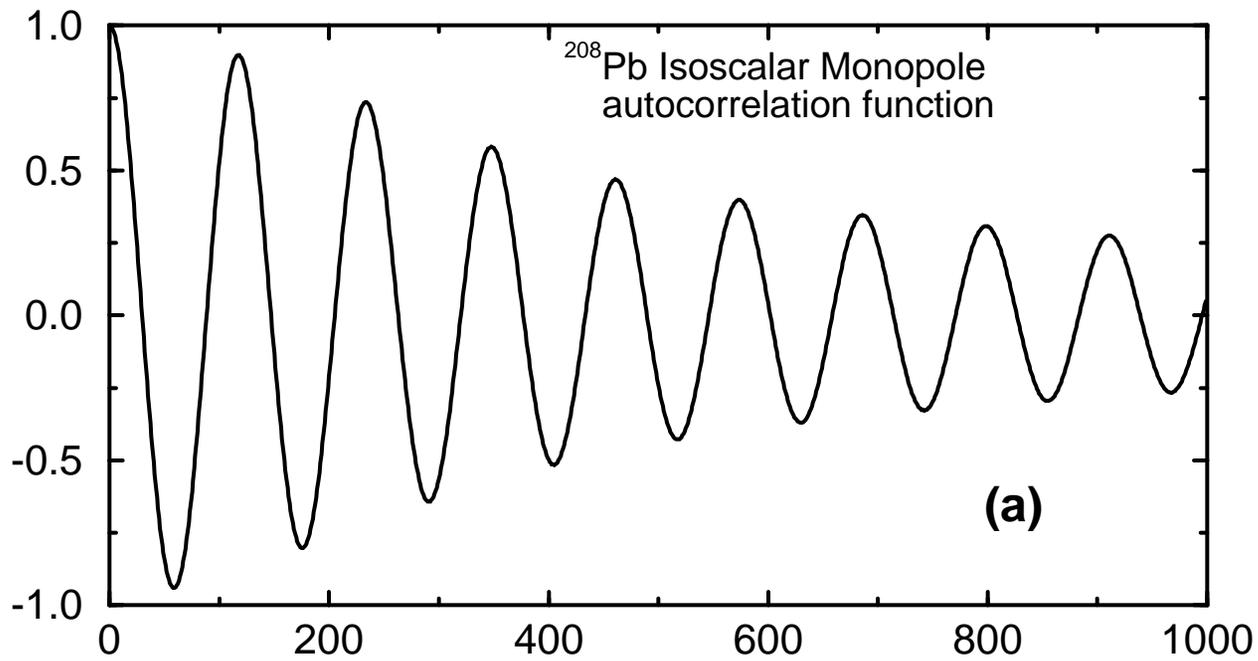
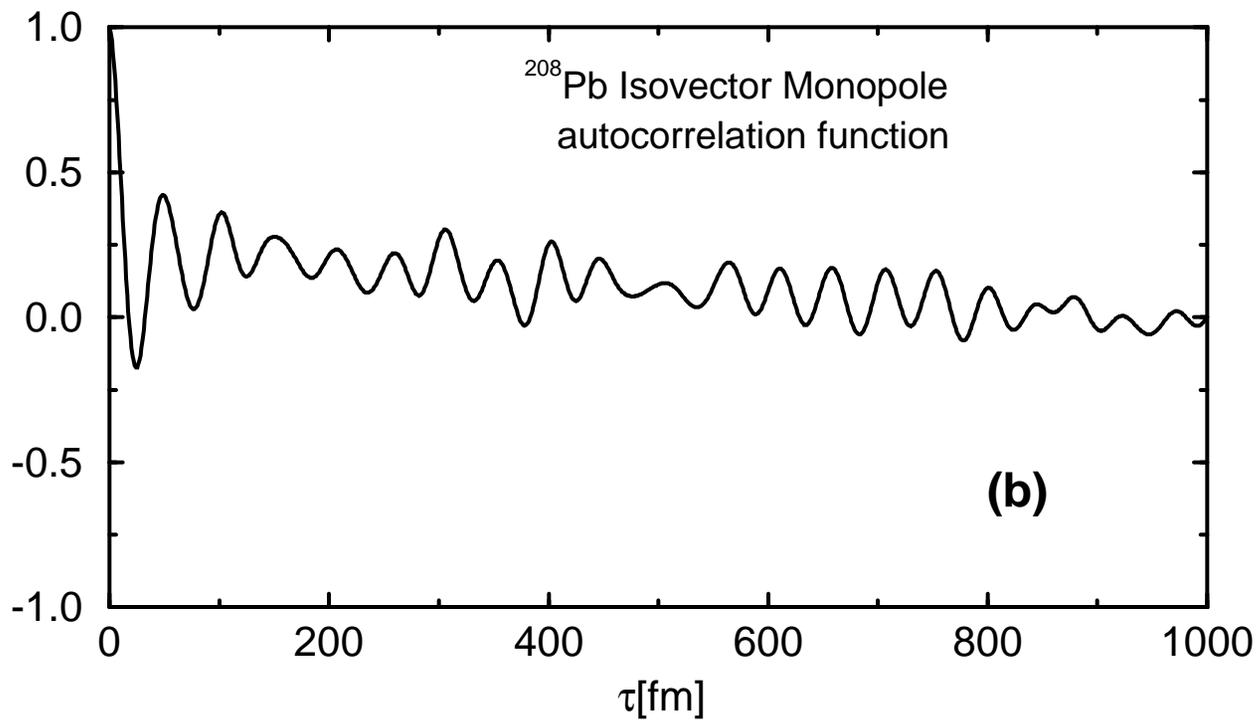
\item[Fig.~4] Autocorrelation functions for isoscalar (a)
and isovector (b) monopole oscillations in $^{208}$Pb.

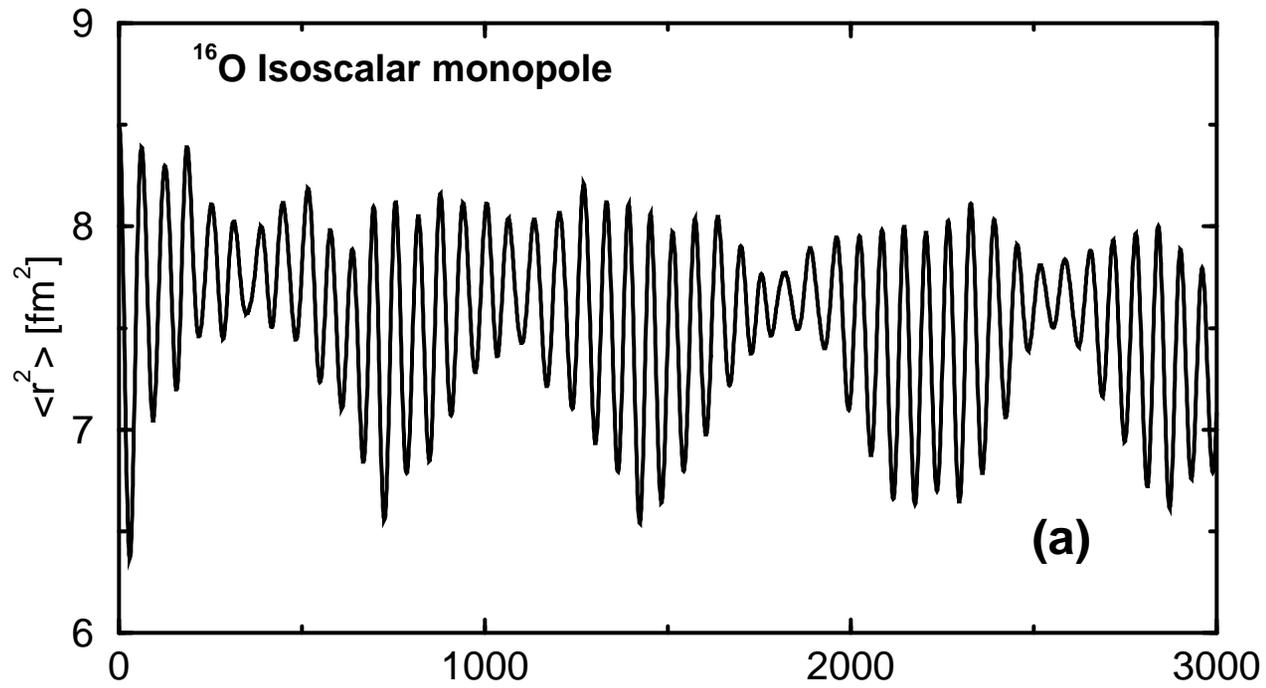
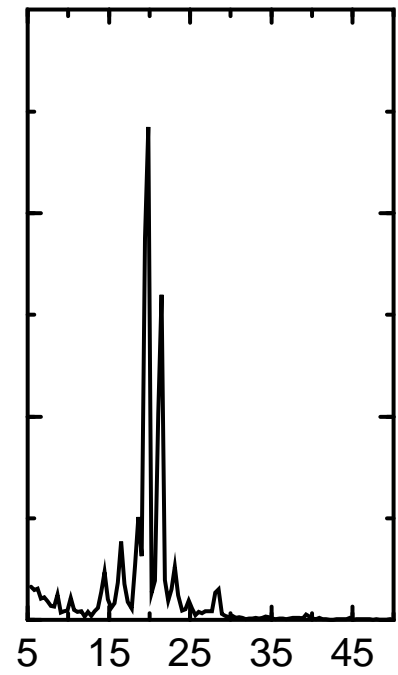
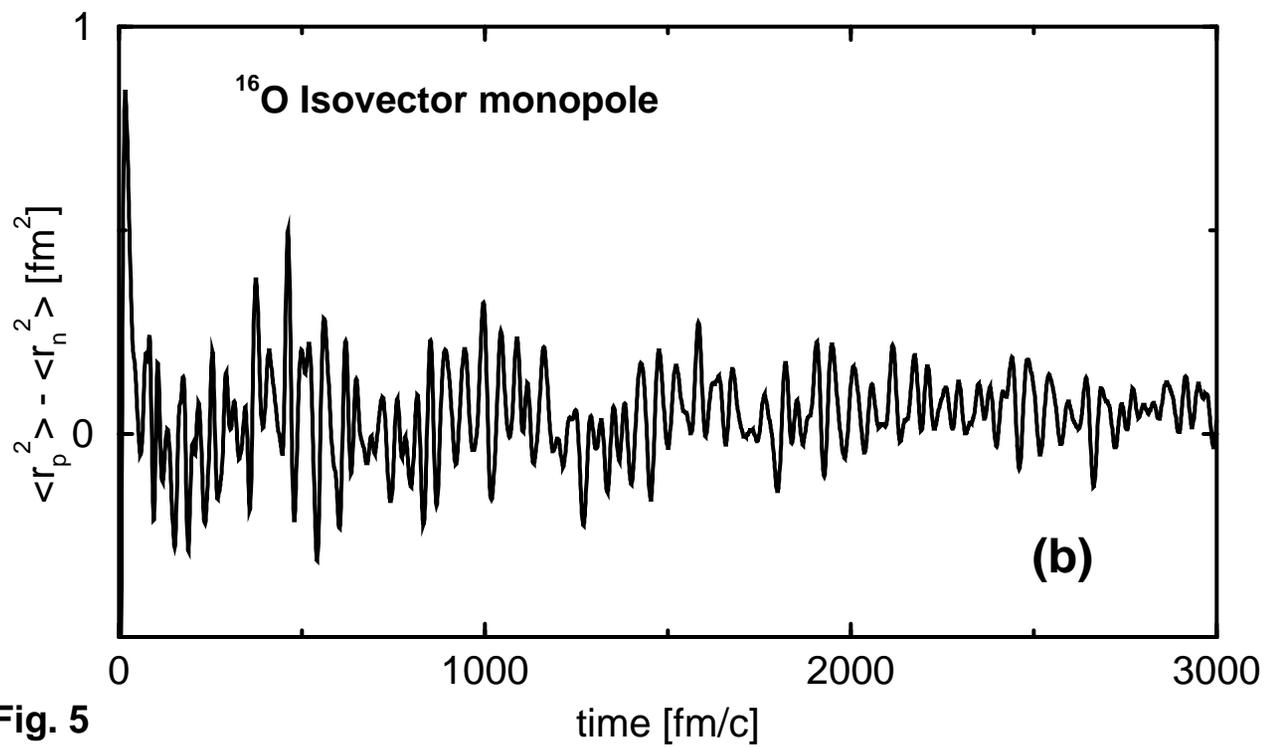
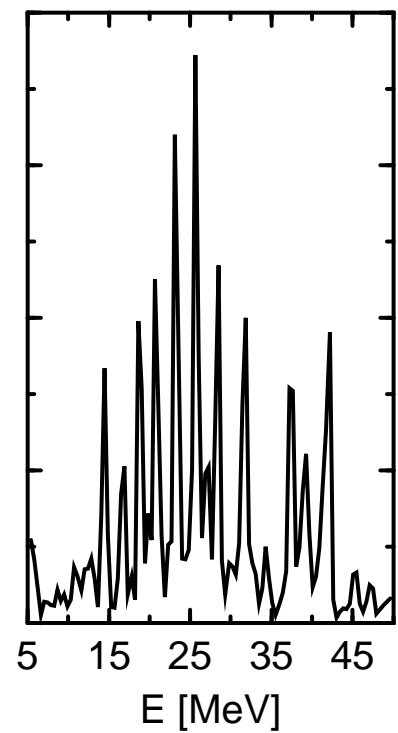
\item[Fig.~5] Time-dependent isoscalar (a) and
isovector (b) monopole moments and the Fourier
spectra for $^{16}$O. The effective force is NL1.

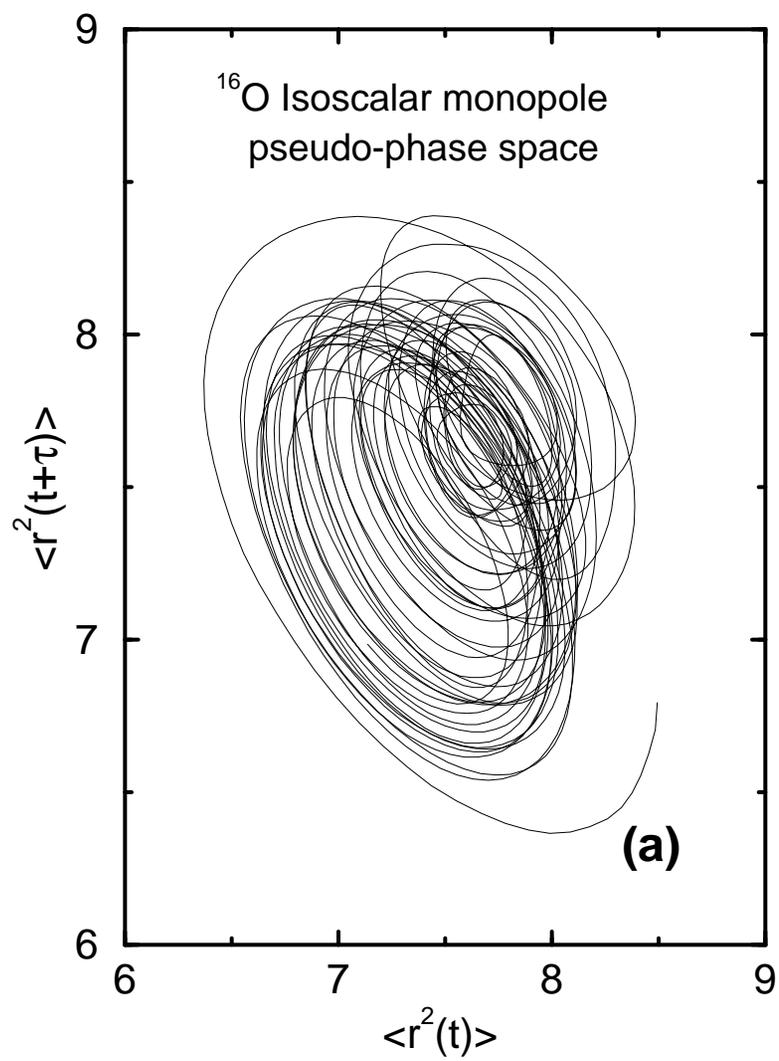
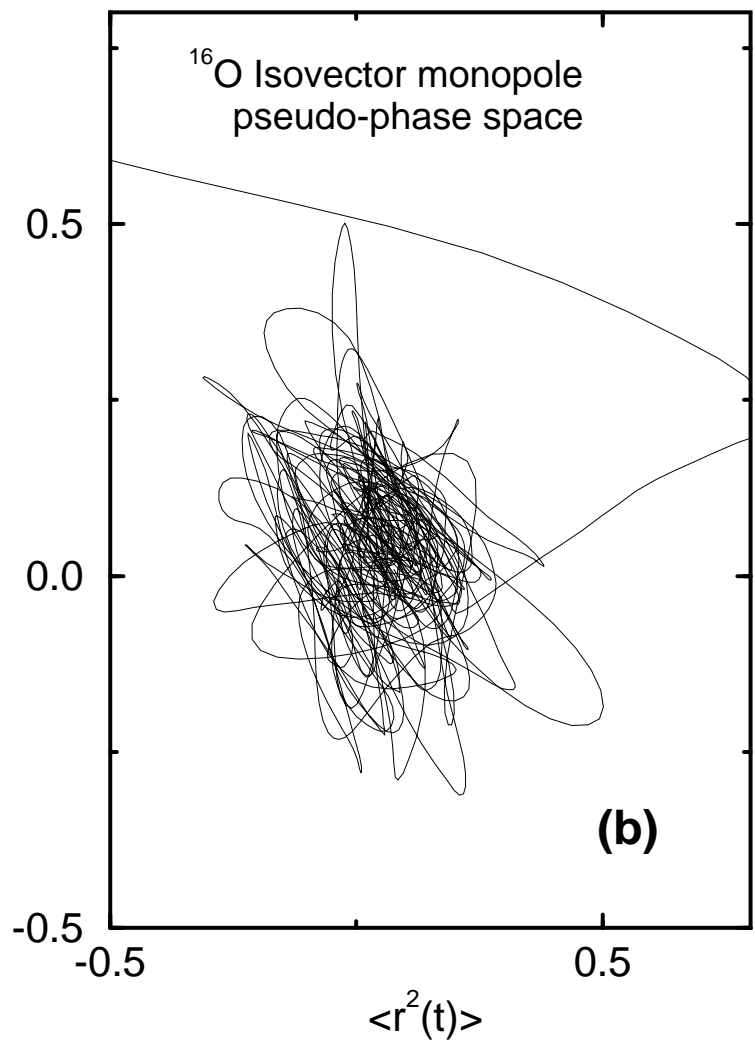
\item[Fig.~6] Two-dimensional pseudo-phase space for
isoscalar (a) and isovector (b) monopole oscillations
in $^{16}$O.

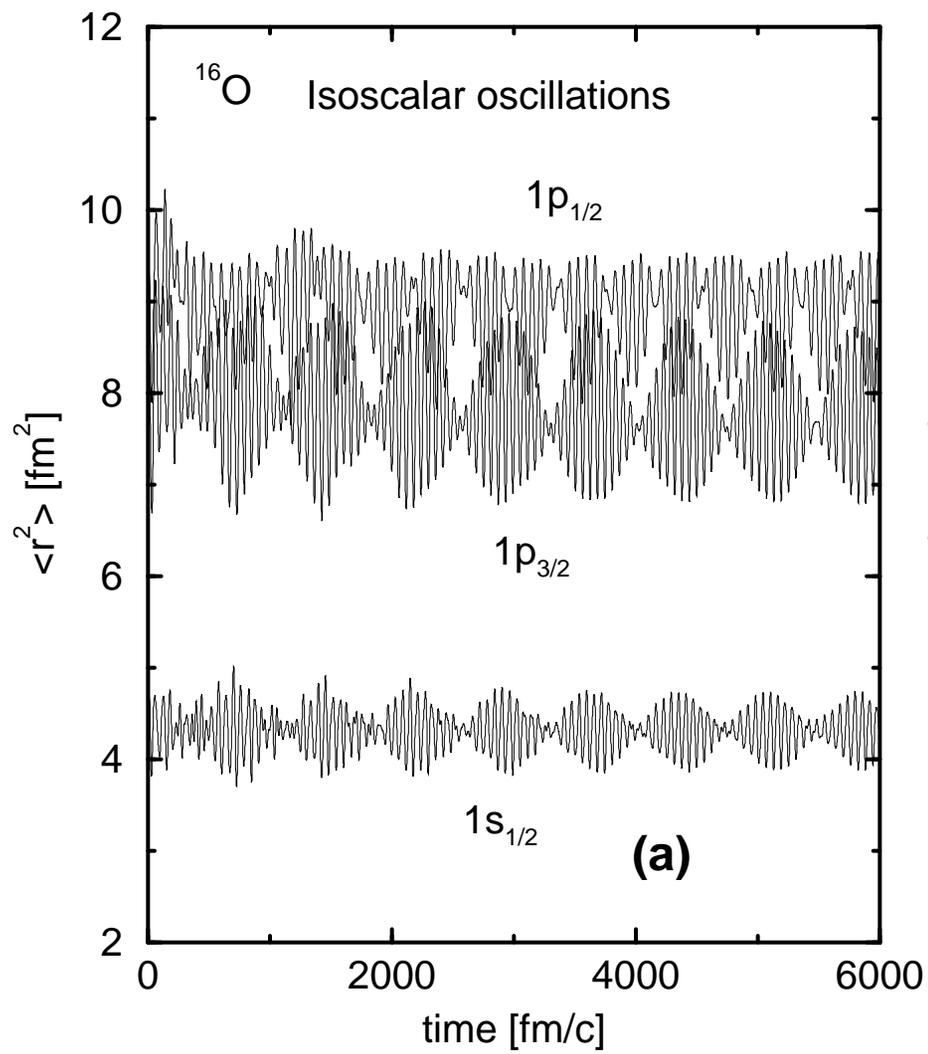
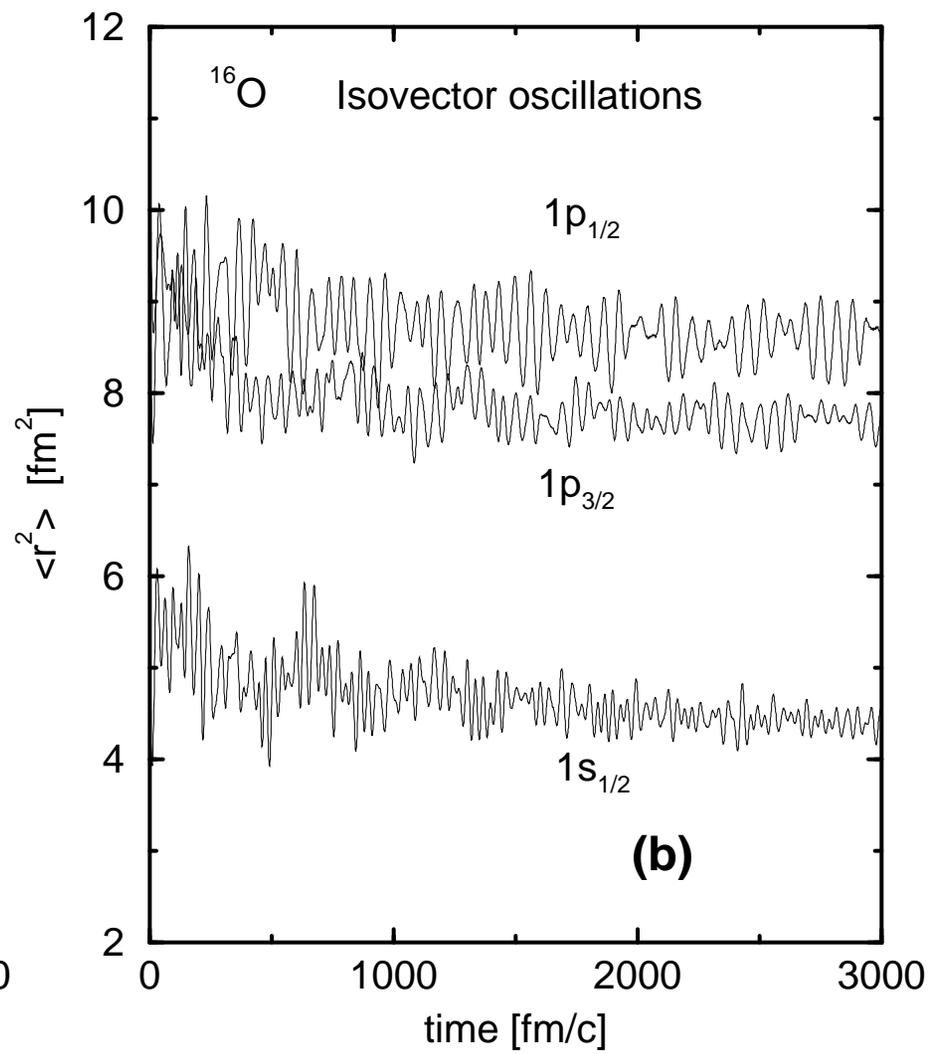
\item[Fig.~7] Time-dependent expectation values $<r^2>$ 
for the three single neutron spherical shell model 
states $1s~1/2$, $1p~3/2$, and $1p~1/2$ in $^{16}$O.
(a) corresponds to the isoscalar oscillations of the 
collective coordinate. In panel (b) results for the 
isovector collective mode are shown.

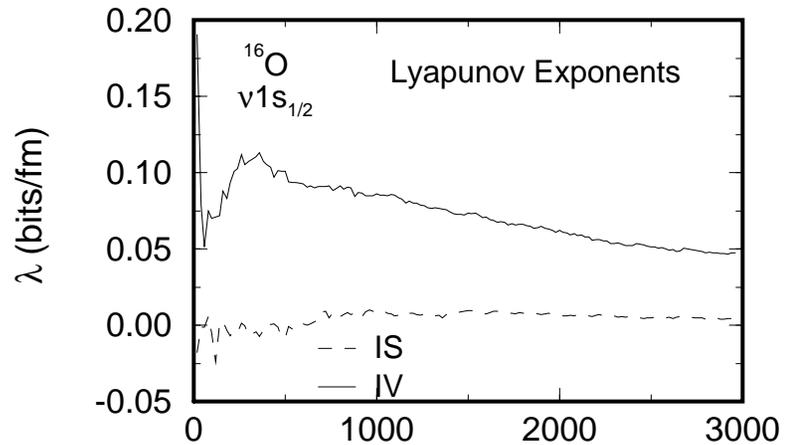
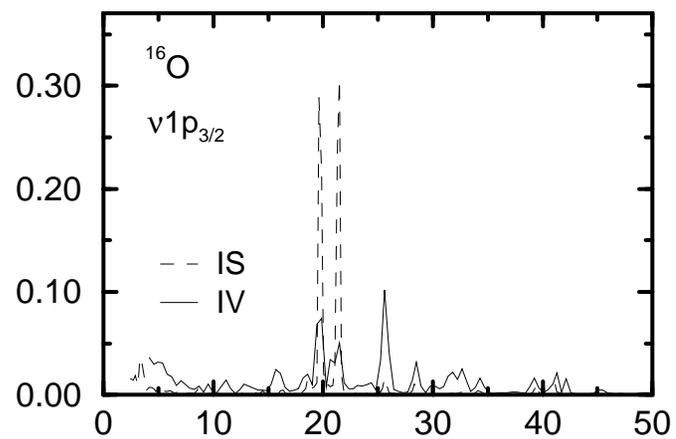
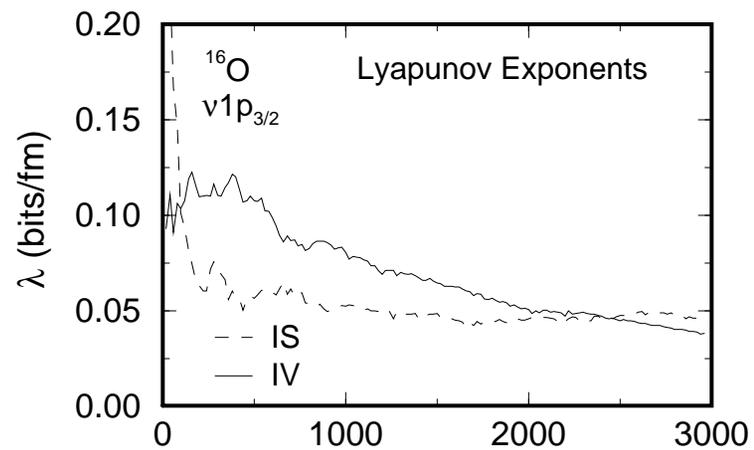
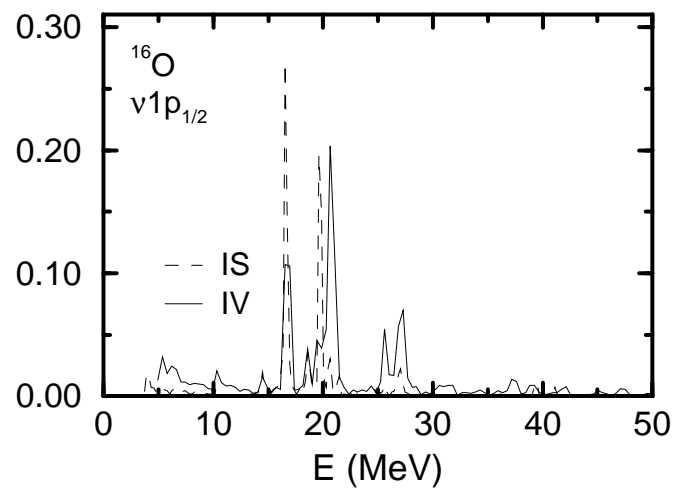
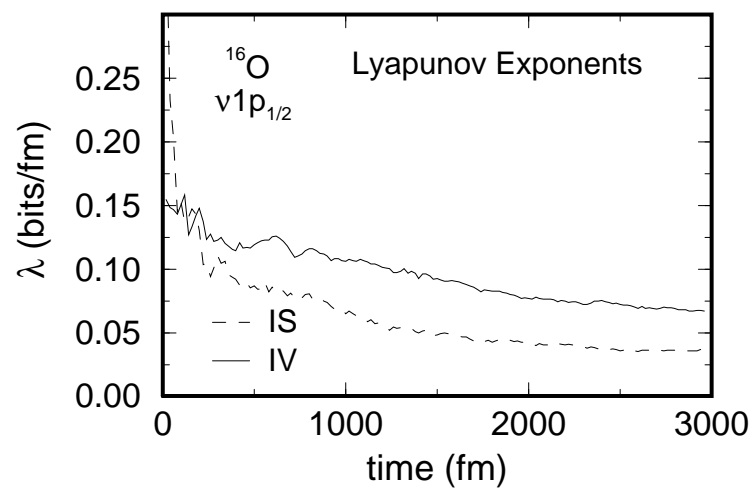
\item[Fig.~8] Fourier spectra and the largest 
Lyapunov exponents for the time-dependent 
expectation values $<r^2>$ in the 
three single neutron spherical shell model
states $1s~1/2$, $1p~3/2$, and $1p~1/2$ in $^{16}$O.
Dashed lines indicate the isoscalar collective monopole mode, 
solid lines correspond to isovector collective oscillations.

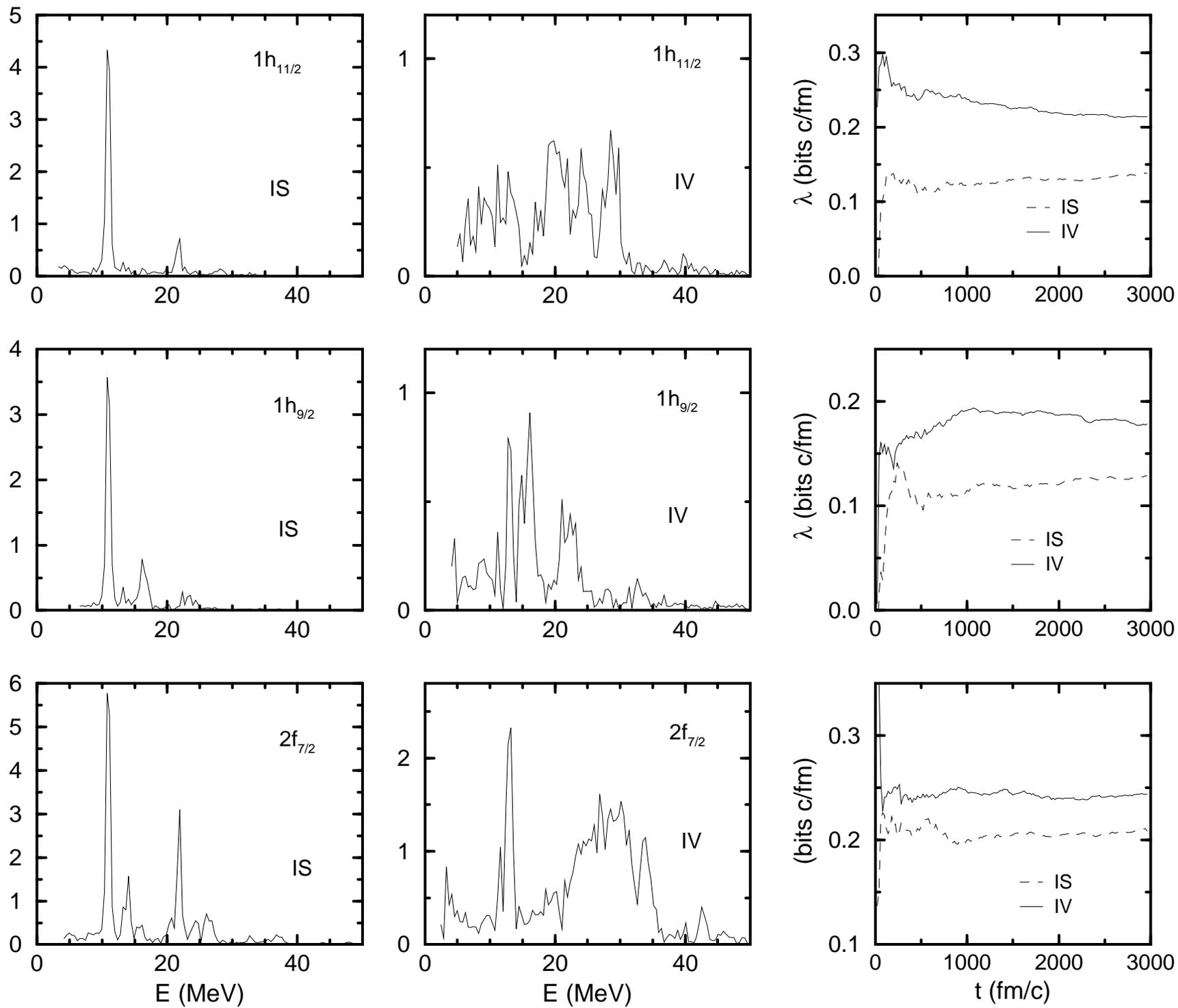
\item[Fig.~9] Fourier spectra and the largest 
Lyapunov exponents for the time-dependent 
expectation values $<r^2>$ in the 
$1h~11/2$, $1h~9/2$, and $2f~7/2$ single neutron
spherical shell model states in $^{208}$Pb.
Results that correspond to the isoscalar collective
monopole mode are shown on the left, those for the isovector
mode are in the center. Lyapunov exponents are displayed 
on right hand side panels.
\end{description}
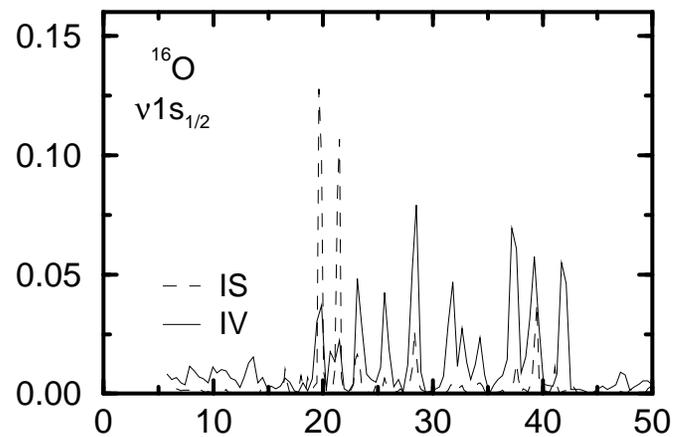
\end{document}